\documentstyle{article}

\def \s{{\rm \;sn}}
\def \c{{\rm \;cn}}
\def \d{{\rm \;dn}}

\def \ch{{\rm \;cosh}}
\def \sh{{\rm \;sinh}}
\def \tan{{\rm \;tan}}
\def \cot{{\rm \;cot}}
\def \sin{{\rm \;sin}}
\def \cos{{\rm \;cos}}

\def \2u{{\frac{\displaystyle u}{\displaystyle 2}}}
\def \id {{\rm \;\cdot id}}

\newcommand {\bt}   {\otimes}
\newcommand {\rw}   {\rightarrow}
\newcommand {\lrw}  {\leftrightarrow}
\newcommand {\Lrw}  {\Leftrightarrow}

\newcommand {\w}    {\omega}
\newcommand {\al}   {\alpha}
\newcommand {\la}   {\lambda}
\newcommand {\ep}   {\epsilon}
\newcommand {\sg}   {\sigma}

\newcommand {\cH}   {{\cal H}}
\newcommand {\cG}   {{\cal G}}
\newcommand {\dps}  {\displaystyle}
\newcommand {\st}   {\stackrel}

\newcommand{\bea}{\begin{array}}   
\newcommand{\eea}{\end{array}}     
\newcommand{\ba}{\begin{eqnarray}}   
\newcommand{\ea}{\end{eqnarray}}     
\newcommand{\be }{\begin{equation}}   
\newcommand{\ee }{\end{equation}}     
\newcommand{\non}{\nonumber}

\newcommand {\rf}[1] {(\ref{#1})}
\newcommand {\dfr}[2]{\frac{\displaystyle {#1}}{\displaystyle {#2}}}
\newcommand {\lbib}[2] { {#1} {\bf #2} }

\newcommand  {\Rmx}    {$R$-matrix{\hspace{5pt}}}
\newcommand  {\Rmxs}   {$R$-matrices{\hspace{5pt}}}

\newcommand{\CMP}{{\em  Commun. Math. Phys.}}

\newcommand{\JPA}{{\em  J. Phys. A: Math Gen.}}
\newcommand{\JMP}{{\em  J. Math. Phys.}}
\newcommand{\JPSJ}{{\em  J. Phys. Soc. Japan.}}

\newcommand{\NP}{{\em  Nucl. Phys.}}
\newcommand{\PL}{{\em  Phys. Lett.}}
\newcommand{\PR}{{\em  Phys. Rev.}}

\newcommand{\IJMP}{{\em Int. J. Mod. Phys.}}

\catcode`\@=11

\def\theequation{\thechapter.\arabic{equation}}

\@addtoreset{equation}{chapter}
\@addtoreset{equation}{section}
\def\theequation
{\ifnum\value{section}>0\relax\thesection.\arabic{equation}\relax
\else\ifnum\value{chapter}>0\relax\thesection.\arabic{equation}\relax
         \else\arabic{equation}\fi\fi}

\begin{document}


\begin{center}
{\LARGE\bf Classification of  Solutions to Reflection Equation of
Two-Component Systems  }

\vspace{0.6cm}

{\large\bf Cong-xin Liu$^{\dagger}$, Guo-xing Ju$^{\ddagger,\S}$, 
Shi-kun Wang$^{\dagger,\P}$, Ke Wu$^{\ddagger}$}

{\sl ${\dagger}$  Institute of Applied Mathematics, \break
           Academia Sinica, \break
           Beijing 100080, China
}
\break
{\sl ${\ddagger}$ Institute of Theoretical Physics, \break
        Academia Sinica, \break
        Beijing 100080, China
}
\break
{\sl ${\S}$ Physics Department,  \break
          Henan Normal University,  \break
          Xinxiang, Henan Province 453002, China \footnote{ Mailing address}
}
\break
{\sl ${\P}$ CCAST(World Laboratory), Beijing 100080, China
}
\end{center}

\vskip .5 cm

\begin{abstract}
The symmetries, especially those related to the $R$-transformation, of
the reflection equation(RE)
for two-component systems are analyzed. The classification of
solutions to the RE for  eight-, six- and seven-vertex type $R$-matrices
is given. All solutions can be obtained from those corresponding to the 
standard $R$-matrices by $K$-transformation. For the free-Fermion models,  
the boundary
matrices have property $tr K_+(0)=0$, and the free-Fermion type $R$-matrix with the 
same symmetry as that of Baxter type corresponds to the same form of $K_-$-matrix 
for the Baxter type. We present the Hamiltonians for
the open spin systems connected with our solutions. In particular, the boundary 
Hamiltonian of  
seven-vertex models was obtained with a generalization to the Sklyanin's formalism.
\end{abstract}


\section{Introduction}
In the framework of quantum inverse scattering
method(QISM)\cite{fad1, ks, fad2, kor, bax}, Yang-Baxter equation(YBE)
\begin{equation}
  R_{12}(u)R_{13}(u+v)R_{23}(v)=R_{23}(v)R_{13}(u+v)R_{12}(u)
\label{ybe}
\end{equation}
 is a sufficient condition for the integrability of
systems with periodic boundary condition(BC). Given a solution
$R$-matrix to YBE (\ref{ybe}), we can construct the Lax operator of certain models
at suitable representation of $R$, and hence transfer matrix
$t(u)$. The YBE ensures that $t(u)$ commutes with each other for
different spectrum parameters. So, if we expand $t(u)$ with respect to
spectrum parameter $u$, the coefficients are a  set of
conserved quantities which satisfy Liouville's criterion
of integrability \cite{fad:ham, das}.

However, when considering  systems on a finite interval with
independent boundary conditions at each end, we have to introduce
reflection matrices $K_{\pm}(u)$ to describe such boundary conditions.
Sklyanin assumed that $R$-matrix has the following
symmetries\cite{sk:bound},
\begin{description}
\item[] Regularity: $R(0)\propto P$;
\item[] P-symmetry: $P_{12}R_{12}(u)P_{12}=R_{21}(u)=R_{12}(u)$;
\item[] T-symmetry: $R_{12}^{t_1t_2}=R_{12}(u)$;
\item[] Unitarity:  $R_{12}(u)R_{21}(-u)\propto id$;
\item[] Crossing Unitarity: $R_{12}^{t_1}(u)R_{21}^{t_1}(-u-2\eta)
                             \propto id$,
\end{description}
where $\eta$ is crossing parameter and $P_{12}$ is the permutation
matrix. $t_1,t_2$
denote transposition in space $V_1$ and $V_2$ respectively.
 In order that the BC  are compatible with integrability, the
reflection matrices should obey so-called reflection equations(RE),
or  boundary Yang-Baxter equations(BYBE)\cite{che, sk:bound, sch}
\ba
      R_{12}(u-v)\st{1}{K}_-(u)R_{21}(u+v)\st{2}{K}_-(v)=
            \nonumber \\
      \st{2}{K}_-(v)R_{12}(u+v)\stackrel{1}{K}_-(u)R_{21}(u-v),
\label{re1}
\ea
\ba
      R_{12}(-u+v)\st{1}{K^{t_1}_+}(u)R_{12}(-u-v-2\eta)
      \st{2}{K^{t_2}_+}(v)=
          \nonumber \\
      \st{2}{K_+^{t_2}}(v)R_{12}(-u-v-2\eta)\st{1}{K_+^{t_1}}(u)
      R_{12}(-u+v).
\label{re2}
\ea
where $\st{1}{K}_{\pm}=K_{\pm}\bt 1$, $\st{2}{K}_{\pm}=1\bt K_{\pm}$,
and $R(u)$ satisfies YBE (\ref{ybe}). So for a solution $K_-(u)$ to RE
\rf{re1},
the relation
\be
      K_+(u)=K_-^t(-u-\eta)  \label{iss1}
\ee
gives the solution to equation \rf{re2}. Nevertheless, not all $R$-matrices possesses 
the above-mentioned properties, some generalizations should be made (see e.g.\cite{mez}).
As we will see in section 4, the seven-vertex(7V) models are also beyond
Sklyanin's formalism for their $R$-matrices do not enjoy $T$-symmetry. It
was stated in \cite{fan} that if $R$-matrix has regularity, unitarity, and
crossing unitarity symmetries, but does not have $PT$-invariance, we can propose $K_+(u)$ to 
satisfy the equation:
\ba
      R_{12}(-u+v)\st{1}{K_+}(u)R_{21}(-u-v-2\eta)
      \st{2}{K_+}(v)=
          \non \\
      \st{2}{K_+}(v)R_{12}(-u-v-2\eta)\st{1}{K_+}(u)
      R_{12}(-u+v),
\label{re21}
\ea
and the integrability can be proved as well. There is also an  relation
between the solutions of Eq.\rf{re1} and Eq.\rf{re21}
\be
      K_+(u)=K_-(-u-\eta).  
\label{iss2}
\ee
We will find later that the Baxter type and free-fermion type I
solutions of seven-vertex models are in this case.

Due to the significance of the reflection equations, a lot of work have
been directed to the study of their solutions \cite{mez, ve:z, hou:z,
ina, rcue, chang, jugx}, and the  Hamiltonians of the systems with such
boundary conditions are also constructed.
However, most of those work are based on
the $R$-matrices which are derived directly from the
parametrization of the statistical weight in vertex models.
There in fact exist many kinds of $R$-matrices according to
the classification of eight- and six-vertex type solutions of both the
YBE and the coloured YBE \cite{wad:fm, wsk8, sun}.  It is
tedious to solve the reflection equation for every $R$-matrix.
Fortunately,  all those $R$-matrices of two-component
systems can be obtained by applying particular solution transformation
to standard(or called gauge) ones \cite{wsk8} which satisfy
certain initial conditions(let us call solution
transformation of \Rmx as R-transformation for brevity). The word
"two-component" means that  there exist two states in the systems:
particles and antiparticles
in field theory, spin up and down in spin system, arrow up and down
or right and left in lattice model (see Ref.\cite{wad:7v}).
After a detailed study of reflection equation, we
can show that there  exists a corresponding transformation to the $K$-
matrix (we call it $K$-transformation) to keep RE invariant under
$R$-transformation. Therefore, we only need to concentrate on
$K$-matrices for the standard $R$-matrices.

In this paper we shall focus our attention on solutions to  reflection
equations of two-component systems up to $K$-transformation.
The solutions are divided into three cases, each of corresponds to
eight-vertex(8V), six-vertex(6V) and
seven-vertex(7V) models and will be discussed in section 2, 3, 4,
respectively. For each case, we first analyze the symmetries of the RE,
especially those related to the $R$-transformations, and then find
solutions to
the RE for the Baxter type and free-Fermion type standard $R$-matrices,
respectively. We put
emphasis on new solutions, but for completeness, we also give the
solutions obtained by others.

In section 5, for those solutions given in previous sections, we shall
construct the corresponding local Hamiltonian
of the open spin-chain. The local Hamiltonian means that it only
consists of nearest-neighbor interaction terms. A system with such
Hamiltonian
can be viewed as having coupling with magnetic field on its ends.
Finally we
shall argue that for all boundary conditions to the free-fermion models,
the reflection matrices $K_+(u)$ have property $tr K_+(0)=0$.
This property requires us to derive the Hamiltonian from the second derivative
of the transfer matrix \cite{rcue}. In section 6, we make some remarks and discussions.

\section{ $K_-$-Matrices to Eight-Vertex Model}

In this section, we shall first study the symmetries of RE
and give the $K$-transformation corresponding to $R$-transformation in
\cite{wsk8}. With these discussions, we can concentrate  our attention on the RE for
the standard 8V $R$-matrices, which
are divided into three types: Baxter-type (or XYZ spin-chain\cite{bax}),
Free-Fermion type I (or XY model\cite{cfan, feld:fm}),
and Free-Fermion type II. All $K_-$-matrices associated with these
$R$-matrices are given.

\subsection{ Symmetries of Reflection Equation }

The general eight-vertex $R$-matrix and the corresponding $K_-(u)$
matrix are expressed in the following forms respectively,
\be
   R(u)=\left( \begin{array}{cccc}
        \omega_1(u)& 0 & 0 &\omega_7(u)  \\
        0  & \omega_2(u) & \omega_5(u) & 0 \\
        0  & \omega_6(u) & \omega_3(u)& 0 \\
        \omega_8(u) & 0 & 0 & \omega_4(u)
       \end{array}\right) \;\;,
\label{8v}
\ee
\be
   K_-(u)=\left( \begin{array}{cc}
      a_1(u) & a_2(u) \\ a_3(u) & a_4(u)
      \end{array}  \right).
\ee
Assuming that $R(u)$ is a solution to YBE (\ref{ybe}), then as studied
in \cite{wsk8},
there are four symmetries for eight-vertex type $R$-matrix (\ref{8v}).
\begin{description}
\item[(R.A) Symmetry of interchanging indices.]  If we
 exchange the elements of $R(u)$ as
 $\w_1(u)\lrw\w_4(u)$, $\w_5(u)\lrw\w_6(u)$ 
 or $\w_2(u)\lrw\w_3(u) $, $ \w_7(u)\lrw\w_8(u)$,
 the new matrix also satisfies YBE (\ref{ybe}).

\item[(R.B) The scaling symmetry.] Multiplication of $R(u)$
by an arbitrary function $f(u)$ is still a solution to YBE (\ref{ybe}).

\item[(R.C) Symmetry of spectral parameter.] If we take
 a new spectral parameter $\bar u=\lambda u$ ,
 where $\lambda$ is an constant complex number, the new matrix
 $R(\bar u)$ is still a solution to YBE (\ref{ybe}).

\item[(R.D) Symmetry of weight functions.]  If we replace
weight functions $\w_7(u)$, $\w_8(u)$ by the new ones
\be
     {\bar\w}_7(u)=s^{-1}w_7(u),  \, {\bar\w}_8(u)=s w_8(u),
  \label{w78}
\ee
where $s$ is a non-zero complex constant, the new matrix
is still a solution to YBE (\ref{ybe}).
\end{description}

The symmetries (R.A)-(R.D) are called solution transformations
(or $R$-transformation) of 8V type solution of YBE (\ref{ybe}).

It is convenient for later discussion to use such notation as follows
$$
\bea{l}
   \w_i(u-v)=u_i,\; \w_i(u+v)=v_i,    \\
   a_i(u)=x_i, \;   a_i(v)=y_i.
\eea
$$
Substituting the matrices $R$ and $K_-$ into the reflection equation
(\ref{re1}),
we get sixteen component equations, which are divided into groups
according to  symmetries of the indices:

$$
\bea{lr}
   \left\{
      \bea{l}
       (u_7v_8-u_8v_7)x_1y_4-u_8v_2x_2y_2+u_7v_3x_3y_3
         +u_4v_6(x_2y_3-x_3y_2)=0, \\
       (u_7v_8-u_8v_7)x_4y_1-u_8v_2x_2y_2+u_7v_3x_3y_3
         +u_1v_5(x_2y_3-x_3y_2)=0,
      \eea
    \right.
   &(A.1)  \\
   \\
   \left\{
      \bea{l}
   (u_2v_3-u_3v_2)x_1y_4+u_2v_8x_2y_2-u_3v_7x_3y_3
         +u_5v_1(x_3y_2-x_2y_3)=0,     \\
   (u_2v_3-u_3v_2)x_4y_1+u_2v_8x_2y_2-u_3v_7x_3y_3
         +u_6v_4(x_3y_2-x_2y_3)=0,
      \eea
   \right.
   &(A.2)  \\
   \\
   \left\{
      \bea{l}
         u_7v_1x_1y_1-u_7v_4x_4y_4-u_1v_7x_1y_4+u_4v_7x_4y_1
            +(u_4-u_1)v_2x_2y_2+u_7(v_5-v_6)x_3y_2=0, \\
         u_8v_1x_1y_1-u_8v_4x_4y_4-u_1v_8x_1y_4+u_4v_8x_4y_1
            +(u_4-u_1)v_3x_3y_3+u_8(v_5-v_6)x_2y_3=0,
      \eea
   \right.
   &(A.3)  \\
   \\
   \left\{
      \bea{l}
         u_3v_6x_1y_1-u_3v_5x_4y_4-u_6v_3x_1y_4+u_5v_3x_4y_1
            +(u_5-u_6)v_8x_2y_2+u_3(v_4-v_1)x_3y_2=0, \\
         u_2v_6x_1y_1-u_2v_5x_4y_4-u_6v_2x_1y_4+u_5v_2x_4y_1
            +(u_5-u_6)v_7x_3y_3+u_2(v_4-v_1)x_2y_3=0,
      \eea
   \right.
   &(A.4)  \\
   \\
   \left\{
      \bea{l}
        (u_1v_1-u_3v_2)x_1y_2+(u_7v_8-u_5v_5)x_4y_2-u_5v_1x_2y_1
            +u_1v_5x_2y_4-u_3v_7x_3y_1+u_7v_3x_3y_4=0,  \\
        (u_1v_1-u_2v_3)x_1y_3+(u_8v_7-u_5v_5)x_4y_3-u_5v_1x_3y_1
            +u_1v_5x_3y_4-u_2v_8x_2y_1+u_8v_2x_2y_4=0,  \\
        (u_4v_4-u_3v_2)x_4y_2+(u_7v_8-u_6v_6)x_1y_2-u_6v_4x_2y_4
            +u_4v_6x_2y_1-u_3v_7x_3y_4+u_7v_3x_3y_1=0,  \\
        (u_4v_4-u_2v_3)x_4y_3+(u_8v_7-u_6v_6)x_1y_3-u_6v_4x_3y_4
            +u_4v_6x_3y_1-u_2v_8x_2y_4+u_8v_2x_2y_1=0,
      \eea
   \right.
   &(A.5)  \\
   \\
   \left\{
      \bea{l}
        u_6v_2x_1y_2-u_1v_7x_1y_3+u_2v_5x_4y_2-u_7v_4x_4y_3
           +(u_2v_1-u_1v_2)x_2y_1+(u_6v_7-u_7v_6)x_3y_1=0,  \\
        u_6v_3x_1y_3-u_1v_8x_1y_2+u_2v_5x_4y_3-u_8v_4x_4y_2
           +(u_3v_1-u_1v_3)x_3y_1+(u_6v_8-u_8v_6)x_2y_1=0,  \\
        u_5v_2x_4y_2-u_4v_7x_4y_3+u_2v_6x_1y_2-u_7v_1x_1y_3
           +(u_2v_4-u_4v_2)x_2y_4+(u_5v_7-u_7v_5)x_3y_4=0,  \\
        u_5v_3x_4y_3-u_4v_8x_4y_2+u_3v_6x_1y_3-u_8v_1x_1y_2
           +(u_3v_4-u_4v_3)x_3y_4+(u_5v_8-u_8v_5)x_2y_4=0.
      \eea
   \right.
   &(A.6)
\eea
$$

After a careful study of the above equations, we find that
if one apply the following transformations to $K_-(u)$ under the
transformations (R.A)-(R.D), the system of equations (A) keeps
invariant:

\begin{description}
\item[(K.A) The symmetry of interchanging indices.]
This symmetry will be discussed for each type of $R$-matrix later.

\item[(K.B) The scalar symmetry.]  If we multiply
$K_-(u)$ by an arbitrary function $g(u)$, the new matrix $g(u)K_-(u)$
is still a solution to RE. On the other hand,
all the \Rmxs  up to an arbitrary scalar function have
the same reflection matrix.

\item[(K.C) The symmetry of spectral parameters.] If we take a new
spectral parameter
$\bar u=\lambda u$ , where $\lambda$ is any constant, the new matrix
 $K(\bar u)$ also satisfies RE for $R(\bar u)$.

\item[(K.D) The symmetry of weight function.]
If applying the transformation (R.D) to $R(u)$,  we can make a
corresponding $K$-transformation on $K_-(u)$:
\be
     {\bar a}_3(u)=\sqrt{s}a_3(u),
     {\bar a}_2(u)=\sqrt{s}^{-1}a_2(u) \label{k23}
\ee
keeping $a_1(u), a_4(u)$  unchanged. The new $K_-(u)$ matrix
is a also solution to RE for new $R$-matrix.
\end{description}

Considering  the above symmetries for $R$-matrix and $K_-$-matrix, we
can concentrate on the
standard \Rmx with the restrictions\cite{wsk8}
\be
    \w_5(u)=\w_6(u)=1,   \,  \w_7(u)=\w_8(u),
 \label{w58}
\ee
and initial condition
\be
    R_{12}(0)=P_{12}.
\ee

Note that from condition \rf{w58}, we only need consider
$R$-transformation (R.A) of interchanging indices $\w_1\lrw\w_4$ and
$\w_2\lrw\w_3$ hereafter.
All $R$-matrices are classified into two classes: Baxter type and
free-fermion type, according to whether or not the elements of
 $R$-matrix satisfy the free-fermion condition
\cite{cfan, bazh, wad:fm}
\be
\w_1(u)\w_4(u)+\w_2(u)\w_3(u)-\w_5(u)\w_6(u)-\w_7(u)\w_8(u)=0.
\ee
The RE corresponding
to these two kinds of $R$-matrices has very different properties.
We shall discuss solutions to RE for these gauge $R$-matrices
respectively.

\subsection{Baxter Type}

The gauge $R$-matrix of Baxter-type was first derived by Baxter
\cite{bax}, it
has the following paramatrization,
\be
  \left\{
    \bea{l}
       \w_1(u)=\w_4(u)=\s (u+h)/\s h, \\
       \w_2(u)=\w_3(u)=\s u/\s h,      \\
       \w_5(u)=\w_6(u)=1,     \\
       \w_7(u)=\w_8(u)=k \s u\s(u+h),
    \eea
  \right.
 \label{b8r}
\ee
where $\s u,\c u,\d u$ are Jacobi elliptic functions of modulus $k$.
It is a high-symmetric one with $\w_1(u)=\w_4(u)$, $w_2(u)=w_3(u)$,
so the transformation of interchanging indices (R.A) has no effect
in this case.

The $K_-$-matrices in this case have been widely discussed in
\cite{ve:z, hou:z, ina, jugx}. For completeness, we list here main
results.
The most general one was given in \cite{ina, jugx} as follows
\begin{equation}
 K_-(u)=\left(\begin{array}{cccc}
      \nu \s(\al-u)   &
      \mu\s(2u)\frac{ \lambda(1-k\s^2 u)+1+k\s^2 u}{1-k^2\s^2\al\s^2 u}
\\
      \mu\s(2u)\frac{ \lambda(1-k\s^2 u)-1-k\s^2 u}{1-k^2\s^2\al\s^2 u}
&
      \nu \s(\al+u)
              \end{array}
     \right),
  \label{b8g}
\end{equation}
where $\mu, \nu, \lambda, \al$ are free parameters,
and the other special solutions can be obtained by setting
these parameters to take special values.

\subsection{Free-Fermion Type I }

The \Rmx of free fermion type-I is less symmetric than that of Baxter
type. In this case, $w_2(u)=w_3(u)$, but $w_1(u)\not=w_4(u)$.
The reflection equation is equivalent to five systems of equations:
\vskip 2ex
$
   \left\{
      \bea{l}
         u_2v_7(x_2y_2-x_3y_3)+u_5v_1(x_3y_2-x_2y_3)=0, \\
         u_2v_7(x_2y_2-x_3y_3)+u_5v_4(x_3y_2-x_2y_3)=0, \\
         u_7v_2(x_2y_2-x_3y_3)+u_1v_5(x_3y_2-x_2y_3)=0, \\
         u_7v_2(x_2y_2-x_3y_3)+u_4v_5(x_3y_2-x_2y_3)=0,
      \eea
    \right.
$
\hfill (B.1)
\vskip  1ex
$
   \left\{
      \bea{l}
        u_7v_1x_1y_1-u_1v_7x_1y_4+u_4v_7x_4y_1-u_7v_4x_4y_4
              +(u_4-u_1)v_2x_2y_2=0,        \\
        u_7v_1x_1y_1-u_1v_7x_1y_4+u_4v_7x_4y_1-u_7v_4x_4y_4
              +(u_4-u_1)v_2x_3y_3=0,
      \eea
   \right.
$
   \hfill (B.2)
\vskip  1ex
$
   \left\{
      \bea{l}
        u_2v_5(x_1y_1-x_4y_4)+u_5v_2(x_4y_1-x_1y_4)
              +u_2(v_4-v_1)x_2y_3=0,        \\
        u_2v_5(x_1y_1-x_4y_4)+u_5v_2(x_4y_1-x_1y_4)
              +u_2(v_4-v_1)x_3y_2=0,
      \eea
   \right.
$
\hfill (B.3)
\vskip  1ex
$
   \left\{
      \bea{l}
        u_5v_2x_1y_2-u_1v_7x_1y_3+u_2v_5x_4y_2-u_7v_4x_4y_3
              +(u_2v_1-u_1v_2)x_2y_1+(u_5v_7-u_7v_5)x_3y_1=0, \\
        u_5v_2x_1y_3-u_1v_7x_1y_2+u_2v_5x_4y_3-u_7v_4x_4y_2
              +(u_2v_1-u_1v_2)x_3y_1+(u_5v_7-u_7v_5)x_2y_1=0, \\
        u_5v_2x_4y_2-u_4v_7x_4y_3+u_2v_5x_1y_2-u_7v_1x_1y_3
              +(u_2v_4-u_4v_2)x_2y_4+(u_5v_7-u_7v_5)x_3y_4=0, \\
        u_5v_2x_4y_3-u_4v_7x_4y_2+u_2v_5x_1y_3-u_7v_1x_1y_2
              +(u_2v_4-u_4v_2)x_3y_4+(u_5v_7-u_7v_5)x_2y_4=0,
      \eea
   \right.
$
\hfill (B.4)
\vskip  1ex
$
   \left\{
      \bea{l}
        (u_1v_1-u_2v_2)x_1y_2+(u_7v_7-u_5v_5)x_4y_2-u_5v_1x_2y_1
              +u_1v_5x_2y_4-u_2v_7x_3y_1+u_7v_2x_3y_4=0,      \\
        (u_1v_1-u_2v_2)x_1y_3+(u_7v_7-u_5v_5)x_4y_3-u_5v_1x_3y_1
              +u_1v_5x_3y_4-u_2v_7x_2y_1+u_7v_2x_2y_4=0,      \\
        (u_4v_4-u_2v_2)x_4y_2+(u_7v_7-u_5v_5)x_1y_2-u_5v_4x_2y_4
              +u_4v_5x_2y_1-u_2v_7x_3y_4+u_7v_2x_3y_1=0,      \\
        (u_4v_4-u_2v_2)x_4y_3+(u_7v_7-u_5v_5)x_1y_3-u_5v_4x_3y_4
              +u_4v_5x_3y_1-u_2v_7x_2y_4+u_7v_2x_2y_1=0.
      \eea
   \right.
$
   \hfill (B.5)
\vskip  2ex

There also exist symmetries of interchanging indices. The system of
equations (B) is invariant under exchange of $a_1(u)\lrw a_4(u)$
and $\w_1(u)\lrw \w_4(u)$ or $a_2(u)\lrw a_3(u)$. The gauge $R$-matrix
is \cite{wsk8},
\be
   \left\{
      \bea{l}
          \w_1(u)=\c u+H \s u\d u,        \\
          \w_4(u)=\c u-H \s u\d u ,       \\
          \w_2(u)=\w_3(u)=G \s u \d u,    \\
          \w_5(u)=\w_6(u)=\d u,           \\
          \w_7(u)=\w_8(u)=k \s u \c u,
       \eea
   \right.
   \label{r1f8}
\ee
where $G$, $H$ are arbitrary parameters with
relation $G^2-H^2=1$. Note that in (\ref{r1f8}) we do not take
$\w_5(u)=\w_6(u)=1$ in oder to compare our following discussion with
other's work. We will consider the general $R$-matrix
which has $w_1(u)\not=\w_4(u)$, i.e. $H\not=0$. The case of
$H=0$ is remarked at the end of this subsection.
Now we solve the  RE (B) case by case.

{\it Case 2.3.1: Diagonal solution.}
From (B.1), $a_2(u)\equiv 0 \Lrw a_3(u)\equiv 0$. There are only two
equations to be considered,
\be
   \left\{
      \bea{l}
         u_2v_5(x_1y_1-x_4y_4)+u_5v_2(x_4y_1-x_1y_4)=0,  \\
         u_7v_1x_1y_1-u_1v_7x_1y_4+u_4v_7x_4y_1-u_7v_4x_4y_4=0.
      \eea
   \right.
\ee
Introducing new variable $g(u)=a_1(u)/a_4(u)$
and solving
$g(u)$ from the above equations, we get a  solution
\begin{equation}
K_-(u)=\left( \begin{array}{cc}
\c u\d u\pm i k'\s u & 0 \\
0 & \c u\d u\mp i k' \s u
     \end{array}  \right),
\label{f8d1}
\end{equation}\\
where $k'$ is the complementary modulus of  elliptic function.
Note that the diagonal solution of 8V free-fermion type I  has no free
parameter, which  is different from that of Baxter type.

{\it Case 2.3.2: Skew-diagonal solution.}
 If $a_2(u)\not\equiv 0$, we conclude from (B.1) that
\be
       a_2(u)=\ep a_3(u), \, \ep=\pm 1.
    \label{f23}
\ee
Taking $a_1(u)\equiv 0$,
we get from (B.4)
$$
       x_4(u_2v_5y_2-u_7v_4y_3)=0.
$$
With the help of \rf{f23} and \rf{r1f8}, the above equation calls for
$a_4(u)\equiv 0$.
However, this  is contradictory to $a_2(u)\not\equiv 0$ as seen from
(B.2).
So the skew-diagonal solution
does not exist due to less symmetry of $R$-matrix.

{\it Case 2.3.3: General solution.}
Because $(u_1-u_4)v_2=(v_1-v_4)u_2$, the
following equation is obtained from (B.2), (B.3) and \rf{f23}
     \ba
        u_7v_1x_1y_1-u_1v_7x_1y_4+u_4v_7x_4y_1-u_7v_4x_4y_4 \non \\
        -\ep\{u_2v_5(x_1y_1-x_4y_4)+u_5v_2(x_4y_1-x_1y_4)\}=0.
     \ea
Differentiating the above equation with respect to $v$ and setting
$v=0$,
we can express $a_1(u),a_4(u)$ as
$$
\left\{
 \begin{array}{c}
    a_1(u)=(F(u)\c u\d u-G(u)\s u)p(u)/2,  \\
    a_4(u)=(F(u)\c u\d u+G(u)\s u)p(u)/2,
 \end{array}
\right.
$$
where
\ba
&&F(u)=c_1+\frac{\displaystyle k((1-\epsilon k G)c_1+H c_2)}
         {\displaystyle \epsilon G-k } \s^2u, \\
&&E(u)=c_2+\frac{\displaystyle k((1-\epsilon k G)c_2-k'^2 H c_1)}
         {\displaystyle \epsilon G-k} \s^2u.
\ea
and  $p(u)$ is a meromorphic function to be determined.
Substituting the above
expressions into (B.5), we get
$$
\frac{a_2(u)}{p(u)}=\mu \s u\c u \d u,
$$
and an additional restriction between $c_1$ and $c_2$ from (B.2)
$$
k'^2 c_1^2+c_2^2=2 \mu^2(G-\epsilon k)/k.
$$
Therefore, the most general solution is
\begin{equation}
K_-(u)=\left( \begin{array}{cc}
F(u)\c(u)\d(u)+E(u)\s(u) &  2\mu\s(u)\c(u)\d(u)\\
2\epsilon\,\mu\s(u)\c(u)\d(u) & F(u)\c(u)\d(u)-E(u)\s(u)
     \end{array}
     \right)
     \label{f8g1}.
\ee
The result in \cite{chang} is a specific case of $\mu=1$.
It is also easy to find  that
the diagonal solution \rf{f8d1} can be obtained by setting
$\mu=0$.

{\bf Remark 2.3.1}. There are in fact  various parameterizaions  of free
fermionic 8V $R$-matrix, one of which is given in \cite{bazh}
\begin{equation}
   \left\{
    \bea{l}
       \w_1(u)=1-e(u)e(h_1)e(h_2),   \\
       \w_4(u)=e(u)-e(h_1)e(h_2),    \\
       \w_2(u)=e(h_2)-e(u)e(h_1),    \\
       \w_3(u)=e(h_1)-e(u)e(h_2),    \\
       \w_5(u)=\w_6(u)=\sqrt{e(h_1)\s(h_1)e(h_2)\s(h_2)}
              (1-e(u))/\s(\2u)\\
       \w_7(u)=\w_8(u)=-i\,k\sqrt{e(h_1)\s(h_1)e(h_2)\s(h_2)}
             (1+e(u))\s(\2u),
    \eea
  \right.
    \label{f8e}
\end{equation}
where $h_1$ and $h_2$ are colour parameters, and $e(u)$ is the
elliptic exponential:
$$
e(u)=\c(u)+i \s(u).
$$
If we make transformation (R.B) to \rf{f8e} with factor function
$$
\sqrt{e(h_1)e(h_2)\s h_1\s h_2}\dfr{1-e(u)}{\s u/2}
$$
and set
$$
    h_1=h_2=h,\, u\rw u/2,\,
    G=\dfr{1}{\s h},\,  H=\dfr{\c h}{\s h},
$$
the new $R$-matrix  coincides with
\rf{r1f8}, so our solution includes the diagonal solution given  in
\cite{rcue}.

{\bf Remark 2.3.2.} Let us consider the special case of $H=0$ and
$G=1$. In this case, the $R$-matrix is
\be
   \left\{
      \bea{l}
          \w_1(u)=\w_4(u)=\c u,        \\
    \w_2(u)=\w_3(u)=\s u \d u,    \\
    \w_5(u)=\w_6(u)=\d u,           \\
          \w_7(u)=\w_8(u)=k \s u \c u.
       \eea
   \right.
   \label{r2f8}
\ee
It has the same symmetry as Baxter type. The calculation shows
that this feature is responsible for the fact that both two
$R$-matrices share the same $K_-(u)$ as given in \rf{b8g}.

\subsection{Free-Fermion Type II }

This kind of \Rmx  takes the form,

\begin{equation}
\left\{
   \begin{array}{l}
    \w_1(u)=\w_4(u)=\dfr{\ch(\lambda u)}{\cos(\mu u)},   \\
    \w_2(u)=-\w_3(u)=-\dfr{\sh(\lambda u)}{\cos(\mu u)},    \\
    \w_5(u)=\w_6(u)=1,                \\
    \w_7(u)=\w_8(u)=\tan(\mu u),
   \end{array}
\right.
\label{fmpm2}
\end{equation}
where $\la,\mu$ are parameters. The RE in component
forms is equivalent to the following twelve equations:
\vskip  2ex
$
   \left\{
      \bea{l}
       u_2v_7(x_2y_2+x_3y_3)+u_5v_1(x_3y_2-x_2y_3)=0, \\
       u_7v_2(x_2y_2+x_3y_3)+u_1v_5(x_3y_2-x_2y_3)=0,
   \eea
    \right.
$
   \hfill(C.1)  \vskip  1ex
$
   \left\{
      \bea{l}
          u_7v_1(x_1y_1-x_4y_4)+u_1v_7(x_4y_1-x_1y_4)=0, \\
          u_2v_5(x_1y_1-x_4y_4)+u_5v_2(x_4y_1-x_1y_4)=0,
     \eea
    \right.
$
   \hfill(C.2)  \vskip  1ex
$
   \left\{
      \bea{l}
          u_5v_2x_1y_2-u_1v_7x_1y_3+u_2v_5x_4y_2-u_7v_1x_4y_3
       +(u_2v_1-u_1v_2)x_2y_1+(u_5v_7-u_7v_5)x_3y_1=0, \\
          u_5v_2x_1y_3-u_1v_7x_1y_2+u_2v_5x_4y_3-u_7v_1x_4y_2
       +(u_2v_1-u_1v_2)x_3y_1+(u_5v_7-u_7v_5)x_2y_1=0, \\
          u_5v_2x_4y_2-u_1v_7x_4y_3+u_2v_5x_1y_2-u_7v_1x_1y_3
       +(u_2v_1-u_1v_2)x_2y_4+(u_5v_7-u_7v_5)x_3y_4=0, \\
          u_5v_2x_4y_3-u_1v_7x_4y_2+u_2v_5x_1y_3-u_7v_1x_1y_2
       +(u_2v_1-u_1v_2)x_3y_4+(u_5v_7-u_7v_5)x_2y_4=0,
     \eea
    \right.
$
   \hfill(C.3)  \vskip  1ex
$
   \left\{
      \bea{l}
          (u_1v_1+u_2v_2)x_1y_2+(u_7v_7-u_5v_5)x_4y_2-u_5v_1x_2y_1
       +u_1v_5x_2y_4+u_2v_7x_3y_1-u_7v_2x_3y_4=0,    \\
          (u_1v_1+u_2v_2)x_1y_3+(u_7v_7-u_5v_5)x_4y_3-u_5v_1x_3y_1
       +u_1v_5x_3y_4+u_2v_7x_2y_1-u_7v_2x_2y_4=0,    \\
          (u_1v_1+u_2v_2)x_4y_2+(u_7v_7-u_5v_5)x_1y_2-u_5v_1x_2y_4
       +u_1v_5x_2y_1+u_2v_7x_3y_4-u_7v_2x_3y_1=0,    \\
          (u_1v_1+u_2v_2)x_4y_3+(u_7v_7-u_5v_5)x_1y_3-u_5v_1x_3y_4
       +u_1v_5x_3y_1+u_2v_7x_2y_4-u_7v_2x_2y_1=0.
     \eea
    \right.
$
   \hfill(C.4)  \vskip  2ex

We find that under $R$-transformation of interchanging $\w_2(u)$ and
$\w_3(u)$ in \rf{fmpm2}, one can perform a $K$-transformation as follows

\be
{\bar a}_2(u)=-a_2(u), \;{\rm or} \;\;\; {\bar a}_3(u)=-a_3(u),
\label{ra72}
\ee
to keep the system of  equations (C) invariant.\\

The existence of nontrivial solution implies that there exists relation
$$\dfr{u_7v_2}{u_2v_7}=\dfr{u_1v_5}{u_5v_1},$$
which  requires $\la=\pm i\mu$. Thus we should consider two
different $R$-matrices,
\begin{equation}
\left\{
  \begin{array}{l}
     \w_1(u)=\w_4(u)=1,   \\
     \w_2(u)=-\w_3(u)=i \tan u,    \\
     \w_5(u)=\w_6(u)=1,\\
     \w_7(u)=\w_8(u)=\tan u,
  \end{array}
\right.
\label{fmpm21}
\end{equation}
and
\begin{equation}
\left\{
\begin{array}{l}
\w_1(u)=\w_4(u)=1,   \\
\w_2(u)=-\w_3(u)=-i \tan u,    \\
\w_5(u)=\w_6(u)=1,\\
\w_7(u)=\w_8(u)=\tan u.
\end{array}
\right.
\label{fmpm22}.
\end{equation}
They are in fact related each other by an
exchange $\w_2(u)\lrw\w_3(u)$. Let us give solution $K_{-}(u)$ directly
because the calculation procedure  has nothing new. For $R(u)$ in
\rf{fmpm21},
we have
\begin{equation}
K_-(u)=\left( \begin{array}{cc}
\mu_1(1+\nu_1 \sin 2u) & i \mu_2(1+\nu_2 \cos 2u)\sin 2u  \\
\mu_2(1-\nu_2 \cos 2u)\sin 2u & \mu_1(1-\nu_1 \sin 2u)
     \end{array}  \right),
\end{equation}
while for $R(u)$ in \rf{fmpm22}, using $K$-transformation \rf{ra72},
we have
\begin{equation}
K_-(u)=\left( \begin{array}{cc}
\mu_1(1+\nu_1 \sin 2u) & i \mu_2(1+\nu_2 \cos 2u)\sin 2u  \\
-\mu_2(1-\nu_2 \cos 2u)\sin 2u & \mu_1(1-\nu_1 \sin 2u)
     \end{array}  \right),
\end{equation}
where $\mu_1, \mu_2, \nu_1, \nu_2$ are free parameters.


\section{ $K_-$-Matrix for Six-Vertex Model}

The general six-vertex \Rmx takes the form
\be
   R(u)=\left( \begin{array}{cccc}
        \omega_1(u)& 0 & 0 & 0      \\
        0  & \omega_2(u) & \omega_5(u) & 0 \\
        0  & \omega_6(u) & \omega_3(u)& 0 \\
        0  & 0           & 0          & \omega_4(u)
       \end{array}\right).
\ee

By setting $u_{7,8}=0$ and $v_{7,8}=0$ in Eqs.(A), we  write down the
reflection equations for 6V type $R$-matrix in component forms:
\vskip  2ex
$
   \left\{
      \bea{l}
             (u_4-u_1)v_2x_2y_2=0, \\
             (u_4-u_1)v_3x_3y_3=0,
      \eea
   \right.
$
   \hfill (D.1)  \vskip  1ex
$
   \left\{
      \bea{l}
             u_1v_5(x_2y_3-x_3y_2)=0, \\
             u_4v_6(x_2y_3-x_3y_2)=0,
      \eea
    \right.
$
   \hfill(D.2)\vskip  1ex
$
   \left\{
      \bea{l}
        (u_2v_3-u_3v_2)x_1y_4+u_5v_1(x_3y_2-x_2y_3)=0,     \\
        (u_2v_3-u_3v_2)x_4y_1+u_6v_4(x_3y_2-x_2y_3)=0,
      \eea
   \right.
$
   \hfill(D.3)  \vskip  1ex
$
   \left\{
      \bea{l}
         u_3v_6x_1y_1-u_3v_5x_4y_4-u_6v_3x_1y_4+u_5v_3x_4y_1
              +u_3(v_4-v_1)x_3y_2=0, \\
         u_2v_6x_1y_1-u_2v_5x_4y_4-u_6v_2x_1y_4+u_5v_2x_4y_1
              +u_2(v_4-v_1)x_2y_3=0,
      \eea
   \right.
$
   \hfill(D.4)  \vskip  1ex
$
   \left\{
      \bea{l}
        u_6v_2x_1y_2+u_2v_5x_4y_2+(u_2v_1-u_1v_2)x_2y_1=0,  \\
        u_6v_3x_1y_3+u_3v_5x_4y_3+(u_3v_1-u_1v_3)x_3y_1=0,  \\
        u_5v_2x_4y_2+u_2v_6x_1y_2+(u_2v_4-u_4v_2)x_2y_4=0,  \\
        u_5v_3x_4y_3+u_3v_6x_1y_3+(u_3v_4-u_4v_3)x_3y_4=0,
      \eea
   \right.
$
   \hfill(D.5)   \vskip  1ex
$
   \left\{
      \bea{l}
        (u_1v_1-u_3v_2)x_1y_2-u_5v_5x_4y_2-u_5v_1x_2y_1
            +u_1v_5x_2y_4=0,  \\
        (u_1v_1-u_2v_3)x_1y_3-u_5v_5x_4y_3-u_5v_1x_3y_1
            +u_1v_5x_3y_4=0,  \\
        (u_4v_4-u_3v_2)x_4y_2-u_6v_6x_1y_2-u_6v_4x_2y_4
            +u_4v_6x_2y_1=0,  \\
        (u_4v_4-u_2v_3)x_4y_3-u_6v_6x_1y_3-u_6v_4x_3y_4
            +u_4v_6x_3y_1=0.
      \eea
   \right.
$
   \hfill(D.6)  \vskip  2ex
From Ref.\cite{sun}, we know that the 6V type solutions
of YBE have the same  solution-transformation
as that for 8V type solutions  except for the symmetries of  weight
functions
and  of the interchanging indices related to $\w_7(u)$ and $\w_8(u)$.
Now the two symmetries of weight functions are
\be
\bar\w_2(u)=s\w_2(u), \,  \bar\w_3(u)=s^{-1}\w_3(u)
\label{tr6v2}
\ee
and
\be
\bar\w_5(u)=e^{c u}\w_5(u), \, \bar\w_6(u)=e^{-c u}\w_6(u)
\label{tr6v5}
\ee
where $s, c$ are two nonzero constants. In fact, we find that
transformation
\rf{tr6v2} has  no effect on the system of equations (D), and
if making $K$-transformation
\be
\bar a_1(u)=e^{c u}a_1(u), \, \bar a_4(u)=e^{-c u}a_4(u)
    \label{tr6k}
\ee
the new $K_-(u)$ is still a solution to RE for the
new \Rmx obtained from $R$-transformation \rf{tr6v5}.
Due to these symmetries, we will
consider the gauge $R$-matrices as  in 8V model.
They are also classified into two classes--
the Baxter type
\be
  \left\{
    \bea{l}
       \w_1(u)=\w_4(u)=\dfr{\sin (u+h)}{\sin h}, \\
       \w_2(u)=\w_3(u)=\dfr{\sin u}{\sin h},      \\
       \w_5(u)=\w_6(u)=1,
    \eea
 \right.
 \label{b6r}
\ee
and the free-Fermion type
\be
 \left\{
    \bea{l}
       \w_1(u)=\dfr{\sin (u+h)}{\sin h}, \\
       \w_4(u)=\dfr{\sin (-u+h)}{\sin h}, \\
       \w_2(u)=\w_3(u)=\dfr{\sin u}{\sin h},      \\
       \w_5(u)=\w_6(u)=1.
    \eea
 \right.
 \label{f6r}
\ee

For Baxter-type, the general solution to RE was given in \cite{ve:z}
\begin{equation}
 K_-(u)=\left(\begin{array}{cc}
      \la \sin(\al-u)   &
 \mu\sin(2u) \\
      \nu\sin(2u) &
        \la \sin(\al+u)
              \end{array}
     \right),
  \label{b6g}
\end{equation}
which has four free parameters $\la, \al, \mu$, and $\nu$.

While for free-fermion type, since $\w_1(u)\not=\w_4(u)$,
one can immediately
see  that $a_2(u)\equiv0$ and $a_3(u)\equiv0$ from
(D.1). In other words, the RE for the free-Fermion type 6V models
only has diagonal solution,
\begin{equation}
 K_-(u)=\left(\begin{array}{cc}
      \sin(\al-u)  &   0 \\
      0                &  \sin(\al+u)
             \end{array}
 \right).
  \label{f6g}
\end{equation}
In addition, if setting  $\cos h=0$ in \rf{f6r}, we have
symmetric $R$ matrix of free-Fermion type as follows,
\be
   R(u)=\left( \begin{array}{cccc}
        \cos u &  0  & 0  & 0  \\
        0  & \sin u & 1 & 0 \\
        0  & 1 & \sin u& 0 \\
        0 & 0 & 0 & \cos u
       \end{array}\right).
   \label{f6s}
\ee
Just as discussed in Remark 2.3.2, this $R$-matrix shares the same
$K_-$-matrix in \rf{b6g} with 6V Baxter-type.

So, up to $K$-transformation \rf{tr6k}, we obtain all general solutions
\rf{b6g} and \rf{f6g} to reflection equation in six-vertex case.

\section{ $K_-$-Matrices to Seven-Vertex Model }

If setting $\w_8(u)\equiv0$ in eight-vertex \Rmx (\ref{8v}), we get
seven-vertex one
\be
   R(u)=\left( \begin{array}{cccc}
        \omega_1(u)& 0 & 0 & \w_7(u)      \\
        0  & \omega_2(u) & \omega_5(u) & 0 \\
        0  & \omega_6(u) & \omega_3(u)& 0 \\
        0  & 0           & 0          & \omega_4(u)
       \end{array}\right). \;\;
\ee
The classification of solutions to the coloured 7V-type YBE
is given recently in \cite{yang7}. Due to less symmetries,
the $R$-matrices show a much more different properties from  that of
both
eight-vertex and six-vertex models. At this point,
we expect that the corresponding reflection equation
reveals new features as well.

First of all, let us study symmetries of reflection
equation as do previously  for other cases.
After removing the terms containing $u_8$ and $v_8$ in the system
of equations (A), we find that there still exists $K$-transformation
\rf{k23} under $R$-transformation \rf{w78}(note that $\w_8$ is absent!).

In \cite{yang7}, an  additional relation  $\w_5(u)/\w_6(u)
=e^{c u}$ is given, where c is a constant. When $c\not=0$,
there have only trivial $K_-$-matrices.
The case of $c=0$ or $\w_5(u)=\w_6(u)$
is  further classified  into three different types:
Baxter type, free-fermion type I, II, which will be discussed
in the following sections.

\subsection{Baxter Type }

The parametrization of the \Rmx is as follows
\be
  \left\{
    \bea{l}
       \w_1(u)=\w_4(u)=\dfr{\sin (u+h)}{\sin h}, \\
       \w_2(u)=\w_3(u)=\dfr{\sin u}{\sin h},      \\
       \w_5(u)=\w_6(u)=1, \\
       \w_7(u)=\sin(u+h)\sin u.
    \eea
  \right.
 \label{b7r}
\ee
Substituting (\ref{b7r}) into (A), we solve these equations case by
case.

{\it Case 4.1.1: Diagonal Solution}. It can be seen from (A.1) that
if $a_2(u)\equiv0$ then $a_3(u)\equiv0$ (Note that $u_8=0=v_8$ in (A)).
 We obtain the diagonal solution:
\begin{equation}
 K_-(u)=\left(\begin{array}{cc}
      \sin(\al-u)  &   0 \\
      0                &  \sin(\al+u)
             \end{array}
 \right).
\label{4b}
\end{equation}

{\it Case 4.1.2: Skew-diagonal solution.} Let $a_1(u)\equiv0$, it
requires
$a_4(u)\equiv0$ from (A.4). We only need to consider two equations:
\be
   \left\{
      \bea{l}
       u_7v_3x_3y_3+u_1v_5(x_2y_3-x_3y_2)=0, \\
       u_3v_7x_3y_3+u_5v_1(x_2y_3-x_3y_2)=0.
      \eea
    \right.
\label{eee}
\ee
Solving Eqs.\rf{eee}, we have  two $K_-$-matrices,
$$
\bea{cc}
\left(\begin{array}{cc}
      0 &   1 \\
      0 &   0
      \end{array}
\right) \,\,\,,
&
\left(\begin{array}{cc}
      0     &   \rho(u)       \\
      1    &   0
      \end{array}
\right)\,,
\eea
$$
where $\rho(u)=(\la+\cos 2u)/2$  and $\la$ is a free parameter.

{\it Case 4.1.3: $a_3(u)\equiv0$.} One can get from case 4.1.1 that
$$
a_1(u)=p(u)\sin(\al-u) , \, a_4(u)=p(u)\sin(\al+u)
$$
and find
$$
a_2(u)/p(u)=\mu \sin(2u),
$$
so the $K_-$-matrix is
\begin{equation}
 K_-(u)=\left(\begin{array}{cc}
      \nu\sin(\al-u)  &   \mu \sin(2u)  \\
       0      &     \nu\sin(\al+u)
             \end{array}
        \right).
        \label{b7s}
\end{equation}
where $\mu$, $\nu$ are parameters.

{\it Case 4.1.4.} Combining the results obtained above,  one can
easily write down the general $K$-matrix as follows
\begin{equation}
 K_-(u)=\left(\begin{array}{cc}
      \nu\sin(\al-u)  &    \mu \rho(u)\sin(2u) \\
      \mu\sin(2u)                &  \nu\sin(\al+u)
             \end{array}
 \right). \label{b7g}
\end{equation}

In summary, we can regard \rf{b7s} and \rf{b7g} as
the most general reflection matrices, because
others can be obtained by assigning special values
to free parameters. Furthermore,
comparing (\ref{4b}) and (\ref{b7s}), we see that in
the case of 7V type, $a_2(u)=0$ implies $a_3(u)=0$, but the
reverse does not hold, this is different from the case of 8V type.

\subsection{ Free-Fermion Type I }

This kind of \Rmx reads
\be
  \left\{
    \bea{l}
       \w_1(u)=\dfr{\sin (u+h)}{\sin h}, \\
       \w_4(u)=\dfr{\sin (-u+h)}{\sin h}, \\
       \w_2(u)=\w_3(u)=\dfr{\sin u}{\sin h},      \\
       \w_5(u)=\w_6(u)=1,   \\
       \w_7(u)=\dfr{\sin 2u}{\sin h}.
    \eea
 \right.
 \label{f7r}
\ee
If $\cos h=0$, or $\w_1(u)=\w_4(u)$, we can make
the similar calculation as do for the Baxter-type, for
both  have the same symmetries. The result is almost
the same as that given in  \rf{b7g} and \rf{b7s} but
 with $\rho(u)=(\la+\cos 2u)$.

When $\w_1(u)\not=\w_4(u)$, it force $a_3(u)\equiv0$ from
the second equation of (A.3).
The RE reduces to the following equations
\vskip  2ex
$
   \left\{
      \bea{l}
         u_2v_5(x_1y_1-x_4y_4)+u_5v_2(x_4y_1-x_1y_4)=0, \\
         u_7v_1x_1y_1-u_7v_1x_4y_4+u_4v_7x_4y_1
        -u_1v_7x_1y_4+(u_4-u_1)v_2x_2y_2=0,
      \eea
   \right.
$
   \hfill(E.1)  \vskip  1ex
$
   \left\{
      \bea{l}
        u_5v_2x_1y_2+u_2v_5x_4y_2+(u_2v_1-u_1v_2)x_2y_1=0,  \\
        u_5v_2x_4y_2+u_2v_5x_1y_2+(u_2v_4-u_4v_2)x_2y_4=0   \\
        (u_1v_1-u_3v_2)x_1y_2-u_5v_5x_4y_2-u_5v_1x_2y_1
            +u_1v_5x_2y_4=0,  \\
        (u_4v_4-u_3v_2)x_4y_2-u_5v_5x_1y_2-u_5v_4x_2y_4
            +u_4v_5x_2y_1=0.
      \eea
   \right.
$
   \hfill(E.2)\vskip  2ex

{\it Case 4.2.1: $a_2(u)\equiv0$.} The two equations of (E.1) are
not compatible with each other due to the symmetry between $\w_1(u)$ and
$\w_4(u)$ being broken. Therefore there exists no diagonal solution for
this type of $R$-matrix.

{\it Case 4.2.2: $a_1(u)\equiv0$.} One can deduce from (E.1) that
$a_4(u)\equiv0$ and
$a_2(u)\equiv0$ . This is a trivial case.

{\it Case 4.2.3: General solution.} From the second equation
of (E.1) and (E.2),
one can get the same result as that in Baxter type
$$
a_1(u)=\sin(\al-u),\, a_4(u)=\sin(\al+u),\, a_2(u)=\mu\sin 2u.
$$
Substituting them into the second equation of (E.1), one find
$\mu=\pm1$. So, we have only one solution
\begin{equation}
 K_-(u)=\left(\begin{array}{cc}
      \sin(\al-u)  &   \pm\sin(2u)  \\
       0      &     \sin(\al+u)
             \end{array}
        \right). \label{f71g}
\end{equation}
which also shows that $a_3(u)=0$ does not imply $a_2(u)=0$.

\subsection{ Free-Fermion Type II }

In this case, the elements of the $R$-matrix take the following forms
\be
 \left\{
    \bea{l}
       \w_1(u)=\w_4(u)=\cosh u, \\
       \w_2(u)=-\w_3(u)=\sinh u,      \\
       \w_5(u)=\w_6(u)=1, \\
       \w_7(u)=u.
    \eea
 \right.
\ee
For the sake of brevity, we simply give the result.
Note that the nontriviality requires $a_1(u)=\pm a_4(u)$.
If $a_1(u)=a_4(u)$, we get
\begin{equation}
 K_-(u)=\left(\begin{array}{cc}
      \al  &   \mu \sinh u  \\
       0     &     \al
             \end{array}
        \right),
\end{equation}
while if $a_1(u)=-a_4(u)$, we have
\begin{equation}
 K_-(u)=\left(\begin{array}{cc}
      \al  &   \mu \cosh u  \\
       0     &  -\al
             \end{array}
        \right).
\end{equation}
In addition, according to the discussion in section 2.4,
the solutions keep invariant under exchange of $\w_2\lrw\w_3$
since $a_3(u)\equiv0$.


\section{Construction of Boundary Hamiltonian}

In this section, we will discuss the Hamiltonians for the systems
described by the
$R$-matrices and $K$-matrices obtained in the previous sections.
The 6V (Baxter type and free-fermion type)
and 8V (Baxter type and free-fermion
type I) are included in Sklyanin's formalism. While for 7V models, both
Baxter type and
free-fermion type I  $R$-matrices has only regularity, $P$-symmetry, unitarity and 
crossing-unitarity
symmetries, their $K_+(u)$-matrices are obtained by \rf{re21}.
However, all of these cases has the same definition of transfer matrix
\cite{sk:bound, fan}, and we can construct their Hamiltonians
in a unified way.

If $K_-(0)\propto id $, $trK_+(0)\not=0$, the Hamiltonian for the open
systems is
defined as\cite{sk:bound}
\be
      \cH\equiv\sum_{j=1}^{N-1}\cH_{j,j+1}+\dfr{1}{2} K_-^{-1}(0)\st{1}
{K^{\prime}}_-(0)+\dfr{tr_0\st{0}{K_+}(0)\cH_{N0}}{trK_+(0)},
\label{bh1}
\ee
where two-site Hamiltonian $\cH_{j,j+1}$ is given by
\be
     \cH_{j,j+1}=P_{j,j+1}\dfr{d}{du}R_{j,j+1}(u)\mid_{u=0}
              =\dfr{d}{du}R_{j,j+1}(u)\mid_{u=0}P_{j,j+1}.
     \label{hjj}
\ee
All the Baxter type models in two-component systems belong
to this case.
The boundary Hamiltonian of 6V and 8V Baxter type
can be found in \cite{ve:z, ina, jugx}.

{\it Case 5.1: Baxter type 7V with crossing parameter $\eta=h$.}
From $K_-(u)$ in \rf{b7g} and relation \rf{iss2}, we find that
\be
 K_+(u)=K_-(-u-h; -\al_+, \mu_+, \nu_+, \la_+).
\ee
According to Eqn.\rf{bh1}, the Hamiltonian is
\ba
\cH=\dfr{1}{4\sin h}\sum^{N-1}_{j=1}\cH_{j,j+1}-A_-\sg^z_1
    +B_-\sg^+_1+C_-\sg^-_1                \non \\
    -A_+\sg^z_N+B_+\sg^+_N+C_+\sg^-_N,
\ea
where
\be
 \cH_{j,j+1}=(2+\sin^2 h)\sg^x_j \sg^x_{j+1}+
       (2-\sin^2 h)\sg^y_j \sg^y_{j+1}+
        i \sin^2 h(\sg^x_j\sg^y_{j+1}+\sg^y_j\sg^x_{j+1})
        +2 \cos h\sg^z_j\sg^z_{j+1},
\ee
\be
A_\pm=\dfr{1}{2}\cot \al_\pm, \;\;
B_\pm=\dfr{(1+\la_\pm)\mu_\pm}{2\sin \al_\pm},\;\;
C_\pm=\dfr{\mu_\pm}{\nu_\pm \sin \al_\pm}.
\ee

However, if $tr K_+(0)=0$, just as pointed out in Refs.\cite{ve:z,
rcue}, there will have no well-defined Hamiltonian  from the first
derivative of the transfer matrix as done in \rf{bh1}. But if
\be
     tr_0\st{0}{K}_+(0)\cH_{N0}=A\id,
 \label{ha}
\ee
where $A$ is a constant,
we can still derive the well-defined local Hamiltonian from the second
derivative of
transfer matrix as follows
\ba
     \cH&\equiv&\dfr{t''(0)}{4(C+2A)}=
        \sum_{j=1}^{N-1}\cH_{j,j+1}+\frac{1}{\dps 2} K_-^{-1}(0)\st{1}
        {K'}_-(0)  \label{bh2} \\
     &&+\dfr{1}{2(C+2A)}\{tr_0(\st{0}{K}_+(0)\cG_{N0})+
       2tr_0(\st{0}{K'}_+(0)\cH_{N0})+tr_0(\st{0}{K}_+(0)\cH_{N0}^2)\},
     \non
\ea
where
\ba
     &&C\equiv trK'_+(0), \\
     &&\cG_{j,j+1}\equiv P_{j,j+1}
      \dfr{d^2 R_{j,j+1}(u)}{du^2}{\dps \mid_{u=0}}.
\ea
The following discussions show that all the boundary conditions
corresponding to the
free-Fermion type $R$-matrix belong to this case. We argue that  it is a
common property for all free-Fermion models.

{\it Case 5.2: Free-Fermion type-I 8V with crossing parameter $\eta=I$.}
Here $I$ is the complete elliptic integral of the first kind of
modulus $k$. For general boundary condition described by  $K_-(u)$ in
\rf{f8g1}, we
have
\ba
  K_+(u)&=&K_-^t(-u-I)       \non  \\
        &=&\left(
  \bea{cc}
    k'^2F_+(u)\s u+E_+(u)\c u\d u & 2\mu_+k'^2\s u\c u\d u \\
    2\ep \mu_+k'^2\s u\c u\d u &  k'^2F_+(u)\s u-E_+(u)\c u\d u
  \eea
     \right)
\ea
where
\ba
&&F_+(u)=c_1^+\d^2u+\frac{\displaystyle k((1-\ep k G)c_1^++H c_2^+)}
         {\displaystyle \ep G-k } \c^2u\,,
\\
&&E_+(u)=c_2^+\d^2u+\frac{\displaystyle k((1-\ep k G)c_2^+-k'^2 H
c_1^+)}
         {\displaystyle \epsilon G-k} \c^2u.
\ea
From Eq.\rf{bh2}, the Hamiltonian reads
\be
   \cH=\sum_{j=1}^{N-1}\cH_{j,j+1}+A_-\sg_1^z+B_-(\sg_1^++\ep \sg_1^-)
                             +A_+\sg_N^z+B_+(\sg_N^++\ep \sg_N^-),
\label{ha1}
\ee
where
\be
    \cH_{j,j+1}=\dfr{H}{2}(\sg_j^z+\sg_{j+1}^z)+\dfr{G+k}{2}
           \sg_j^x\sg_{j+1}^x+\dfr{G-k}{2}\sg_j^y\sg_{j+1}^y,
\ee
and
$$
    A_-=c_2^-/c_1^-, \, B_-=2\mu^-/c_1^- ,
$$
$$
    A_+=k'^2(H F_+(0)-E_+(0))/2(k'^2 F_+(0)+H E_+(0)),
$$
$$
    B_+=k'^2(G+\ep k)\mu^+/(k'^2 F_+(0)+H E_+(0)).
$$
For the diagonal $K$-matrix \rf{f8d1}, we have
\be
   \cH=\sum_{j=1}^{N-1}\cH_{j,j+1}
  +\dfr{i k'}{2}(\sg_1^z-\sg_N^z),
\ee
which has been discussed in \cite{rcue} and is a
special case of (\ref{ha1}).

{\it Case 5.3: Symmetric free-Fermion type-I 8V.}
If considering $R$-matrix  \rf{r2f8} and the corresponding
$K$-matrix\rf{b8g}, we get
the following  Hamiltonian
\ba
      \cH&=&\sum_{j=1}^{N-1}(\dfr{1+k}{2}\sg_j^x\sg_{j+1}^x+
          \dfr{1-k}{2}\sg_j^y\sg_{j+1}^y)  \non \\
     &&-A_-\sg_1^z+B_-\sg_1^+ +C_-\sg_1^-
     -A_+\sg_N^z+B_+\sg_N^+ +C_+\sg_N^-
\ea
where
$$
  A_-=\dfr{\c\al_\pm\d\al_\pm}{2\s\al_\pm},\;\;
  A_+=\dfr{\c \al_+}{2\s\al_+\d\al_+}(1-k^2 \s\al_+),
$$
$$
  B_\pm=\dfr{\mu_\pm(\la_\pm+1)}{\s\al_\pm},\;\;
  C_\pm=\dfr{\mu_\pm(\la_\pm-1)}{\s\al_\pm}.
$$

{\it Case 5.4: Free-Fermion type 6V  with $\eta=\pi/2$.}
 For $R$-matrix
in \rf{f6r} and the general $K_-$-matrix \rf{f6g}, if setting
\be
K_+(u)=K^t_-(-u-\pi/2; \pi/2-\al_+-h)
\ee
we have
\ba
      \cH&=&\dfr{1}{\sin h}\sum_{j=1}^{N-1}(\sg_j^x\sg_{j+1}^x+
          \sg_j^y\sg_{j+1}^y+\cos h(\sg_j^z+\sg_{j+1}^z))  \non \\
     &&-\cot \al_-\sg_1^z-\cot \al_+\sg_N^z.
\ea

{\it Case 5.5:  Symmetric free-Fermion type  6V.}
If considering $R(u)$ in \rf{f6s} together with the
general $K_-$-matrix \rf{b6g}, we can set
\be
   K_+(u)=K^t_-(-u-\pi/2; -\al_+, \mu_+, \nu_+),
\ee
thus the Hamiltonian is
\ba
      \cH&=&\sum_{j=1}^{N-1}(\sg_j^x\sg_{j+1}^x+
          \sg_j^y\sg_{j+1}^y)  \non  \\
   &&-A_-\sg_1^z+B_-\sg_1^+ +C_-\sg_1^-
     -A_+\sg_N^z+B_+\sg_N^+ +C_+\sg_N^-
\ea
where
$$
  A_\pm=\cot \al_\pm, \;\; B_\pm=\dfr{2\mu_\pm}{\sin \al_\pm},\;\;
  C_\pm=\dfr{2\nu_\pm}{\sin \al_\pm}.
$$

{\it Case 5.6: Free-Fermion type I 7V with crossing parameter
$\eta=\pi/2$.} From $K_-$-matrix in \rf{f71g} and relation \rf{iss2}, we
get
\be
K_+(u)=K_-(-u-\pi/2; \al_+-h+\pi/2, \mu_+),
\ee
and the Hamiltonian is
\ba
\cH&=&\dfr{1}{2\sin h}\sum_{j=1}^{N-1}\{\cos h(\sg^z_j+\sg^z_{j+1})
     +2 \sg^x_j \sg^x_{j+1}+i(\sg^x_j\sg^y_{j+1}+
        \sg^y_j\sg^x_{j+1})\}   \non \\
   &&-A_-\sg^z_1+B_-\sg^+_1
     -A_+\sg^z_N+B_+\sg^+_N,
\ea
where $$\mu_-,\mu_+=\pm 1,\;\; A_\pm=\cot \al_\pm/2,\;\;
B_\pm=\dfr{\mu_\pm}{\sin \al_\pm}.$$

{\it Case 5.7:  Symmetric free-Fermion type I  7V.}
From $K_-$-matrix  in \rf{b7g} with $\rho(u)=\la+\cos 2u$ and
relation \rf{iss2}, we have
\be
K_+(u)=K_-(-u-\pi/2; -\al_+, \mu_+,\nu_+, \la_+),
\ee
and
\ba
\cH&=&\sum_{j=1}^{N-1}\{\sg^x_j \sg^x_{j+1}
      +\dfr{i}{2}(\sg^x_j\sg^y_{j+1}+\sg^y_j\sg^x_{j+1})\} \non \\
     &&-A_-\sg_1^z+B_-\sg_1^+ +C_-\sg_1^-
       -A_+\sg_N^z+B_+\sg_N^+ +C_+\sg_N^-
\ea
where
$$
 A_\pm=\dfr{\cot \al_\pm}{2},\;\; B_\pm=\dfr{(1+\la_\pm)\mu_\pm}{\nu_\pm\sin
\al_\pm},\;\;
 C_\pm=\dfr{\mu_\pm}{\nu_\pm\sin \al_\pm}.
$$

It should be pointed out that for the  free-Fermion type II of
both 7V and 8V models which have no crossing-unitarity symmetry,
how to prove their integrability and to obtain the corresponding
Hamiltonians in the case of open boundary condition is an open problem.

\section{Remarks and Discussions}

In this paper we find that symmetries play an important role in
solving the reflection equation.
For any non-standard $R$-matrix which is  obtained by applying
$R$-transformation
to the  standard one,  then the  corresponding reflection matrix can be
obtained by  making
$K$-transformation to that for standard $R$-matrix.

Moreover,  all solutions given above indicate that the number
of free parameters appeared in $K_-$-matrix is determined by symmetries
of $R$-matrix. The $R$-matrices with  different forms but the same
symmetries share the same
$K_-(u)$ matrix. The free-Fermion type $R$-matrix with $\w_1(u)=\w_4(u)$
is just in this case. It  has the same form $K_-(u)$ as in Baxter type.
Also we note that, different from that for six- and eight-vertex cases,
the elements $a_2(u), a_3(u)$ of $K_-(u)$ in seven-vertex case have no
interchanging-symmetry resulting from the symmetry between $\w_7(u)$ and
$\w_8(u)$ of $R$-matrix being broken.

It is also interesting to note that while constructing Hamiltonian,
all reflection matrices for
free-Fermion $R$-matrices have property of $tr K_+(0)=0$. We argue that
it is a typical property for all free-Fermion models.
So the local Hamiltonian for such system are obtained from second derivative
of the transfer matrix.

We are sure that our procedure to find solutions of the reflection
equation can be
applied to high-spin models, though the calculation may be
much more involved in this case.
With the solutions given
in this paper, we can use the Bethe ansatz method to study the physical
properties of open spin chains.

Furthermore, recently much attention
has been directed to the Yang-Baxter equation with dynamical parameters
\cite{fel,av}. How to
construct the corresponding reflection equation and to seek its solution
is an open problem. We wish to discuss some related problems using the
method and procedure given in this paper.

\end{document}